\documentclass[aps,pre,showpacs]{revtex4}
\usepackage{graphicx,epsfig}
\begin{document}
\title{Coupled Map Networks as Communication Schemes}
\author{P. Garc\'{\i}a}
\affiliation{Departamento de F\'{\i}sica Aplicada, 
Facultad de Ingenier\'{\i}a,
Universidad Central de Venezuela, Caracas, Venezuela}  
\author{A. Parravano}
\affiliation{Centro de Astrof\'{\i}sica Te\'orica, Facultad de Ciencias,
Universidad de Los Andes, \\Apartado Postal 26 La~Hechicera, M\'erida~5251,
Venezuela.} 
\author{M. G. Cosenza}
\affiliation{Centro de Astrof\'{\i}sica Te\'orica, Facultad de Ciencias,
Universidad de Los Andes, \\Apartado Postal 26 La~Hechicera, M\'erida~5251,
Venezuela.}
\author{J. Jim\'enez}
\affiliation{ Laboratorio de Fen\'onemos no Lineales, Escuela de F\'{\i}sica,
Facultad de Ciencias\\ Universidad Central de Venezuela, Caracas, Venezuela.}
\author{A. Marcano}
\affiliation{ Laboratorio de Fen\'onemos no Lineales, Escuela de F\'{\i}sica,
Facultad de Ciencias\\ Universidad Central de Venezuela, Caracas, Venezuela.}
\date{Accepted in Phys. Rev. E Rapid Communications, 2002}

\begin{abstract}
Networks of chaotic coupled maps are considered as string and language generators.
It is shown that such networks can be used as encrypting systems where the
ciphertext contains information about the evolution of the network and also about
the way to select the plaintext symbols from the string associated to the network
evolution. The secret key provides the network parameters, such as the coupling
strengths.
\end{abstract}
\maketitle

Most languages produce aperiodic messages with finite entropy \cite{shan64}.
Since this property is emblematic of chaotic systems, they are potential
candidates to model simple languages and to design
communication schemes \cite{bap,ba00,mar00,alv99,Hayes,Roy}.    
Most of these models and schemes have considered
the use of only one chaotic dynamics either for masking the message to be sent,
or for transmiting a controlled signal. However, these procedures 
may result in poor
security when used as a communication system \cite{alv00,Perez,Yang}. 
On the other hand, a large number of connected physiological units are
involved in real languages. Thus, it seems interesting to explore the performance
of a network of interacting chaotic elements as a model to generate simple 
languages and as communication schemes.

In this article we study networks of coupled chaotic maps
as generators of strings of symbols,
and investigate their potential use as an encrypting system. 
A coupled map network (CMN) can be defined as
\begin{equation}
x_{t+1}^i=f(x_t^i) + \sum _{j=1}^N \epsilon _{ij}x_t^j   \,,
\label{cml}
\end{equation}
where $x_t^i$ gives the states of the element $i$ ($i=1,...,N$) at discrete time
$t$; $f(x_t^i)$ is a real function describing the local dynamics; $\epsilon_{ij}$
are the coupling strengths among elements in the system; and $N$ is the size of
the network. Coupled map lattices have provided 
fruitful models for the study of
a variety of spatiotemporal processes in spatially distributed systems 
\cite{Chaos}.  

Equation (\ref{cml}) can be written in vector form as
\begin{equation}
\label{vector}
{\bf x}_{t+1}= {\bf{f}}({\bf{x}}_t)+{\bf{E}} {\bf{x}}_t.
\end{equation}
The state vector ${\bf{x}}_t$ possesses $N$ components ${\bf{x}}_t=(x_t^1,
x_t^2, \ldots,x_t^N)$, corresponding to the states of the elements in the network.
The $N \times N$ elements of matrix ${\bf E}$ are $\epsilon_{ij}$, which we
assume in general different among themselves, i.e., the coupling is heterogeneous.

The system Eq. (\ref{vector}) can be used as a string generator and thus it can
produce a sequence of symbols. For simplicity, we shall consider a CMN consisting
of $N=7$ maps. To each CMN state ${\bf x}_t=(x_t^1,x_t^2, ..., x_t^7)$ we can
assign a binary state $(b_t^1\, b_t^2\, b_t^3\, b_t^4\, b_t^5\, b_t^6\, b_t^7)$
by the following rule: $b_t^i=0$ if $x_t^i < x^*$, and $b_t^i=1$ if $x_t^i > x^*$,
where $x^*$ is some threshold value. With a prefixed correspondence rule, each of
the $128$ possible seven-digit binary states $(b_t^1\, b_t^2\, b_t^3\, b_t^4\,
b_t^5\, b_t^6\, b_t^7)$ can be associated to one ASCII symbol $z_k$ among the set
$Z_{128}=\{z_1,z_2,\ldots,z_{128}\}$. We take $x^*=0$. In this
case, each seven-digit binary state corresponds to one of the  $2^7=128$
``Cartesian quadrant" in the seven-dimensional state space of the CMN, enough to
assign an ASCII symbol $z_k$, ($k=1,2, \ldots, 128$) to each ``quadrant".

Let us assume that, starting from any initial condition 
${\bf{x}}_0$, the state vector of the CMN
visits all the ``quadrants'' during its evolution, so that all ASCII symbols in
$Z_{128}$ are generated by the CMN dynamics. If we assign to the state
${\bf{x}}_t$ the ASCII symbol corresponding to the ``quadrant'' where
${\bf{x}}_t$ lies at time $t$, the string
$\alpha=(z_{k_1},z_{k_2},\ldots,z_{k_t},\ldots,z_{k_T})$ of ASCII symbols will be
generated up to time $T$. We denote by $|\alpha|=T$ the length of the string,
i.e., the number of iterations performed on the CMN system up to time $t=T$.

On the other hand, for a given
set of ordered ASCII symbols $\rho=(p_1,p_2,\ldots,p_n)$, a sufficiently
long string $\alpha$ can be expressed as a succession of substrings $\beta_l
\cdot p_l$,
\begin{equation}
\label{segments} \alpha=(\beta_1 \cdot p_1,\beta_2 \cdot p_2,
\ldots,\beta_{l-1} \cdot p_{l-1},\underbrace{\beta_l \cdot p_l}_{\mbox{l
segment}}, \ldots, \beta_n \cdot p_n) \, ;
\end{equation}
where $\beta_1 \cdot p_1$ is the substring begining at $z_{k_1}$ and ending at
the first occurrence of symbol $p_1$, the substring $\beta_2 \cdot p_2$ begins
after $p_1$ and ends at the first occurrence of symbol $p_2$, and so on. For
example, the string
\begin{equation}
\label{ejemp_a}
\alpha=(d,4,\$,R,m,e,>,i,\&,H,+,t,5,v,?,u,K,g,a,i,a,6,l)
\end{equation}
is segmentated by the word ``Rival", i.e., $\rho=(R,i,v,a,l)$ as
\begin{equation}
\label{ejemp_s}
\alpha=(\, \underbrace{d,4,\$}_{\beta_1},\overbrace{R}^{p_1},
\underbrace{m,e,>}_{\beta_2},\overbrace{i}^{p_2},
\underbrace{\&,H,+,t,5}_{\beta_3},\overbrace{v}^{p_3},
\underbrace{?,u,K,g}_{\beta_4},\overbrace{a}^{p_4}
\underbrace{i,a,6}_{\beta_4},\overbrace{l}^{p_5}\,) \, .
\end{equation}

The set of marker symbols $(p_1,p_2,\cdots,p_l,...,p_n)$ univocally determines
how string $\alpha$ is segmented by the rule in Eq. (\ref{segments}).

Any string $\alpha$  resulting from the evolution of the CMN can always be
expressed as a concatenation of segments $\beta_l \cdot p_l$, provided that
there are no symbols forbidden by the dynamics of the CMN, i.e., all the
``quadrants" are visited by the state vector ${\bf{x}}_t$. Let ${\bf{y}}_{l-1}$
be the state of the CMN when the symbol $p_{l-1}$ occurs at the end of substring
$\beta_{l-1} \cdot p_{l-1}$. The next substring $\beta_l \cdot p_l$ depends only
on ${\bf{y}}_{l-1}$ and on the symbol $p_l$. Then, substring $\beta_l \cdot p_l$
can be expresed as
\begin{equation}
\label{gdxp}
\beta_l \cdot p_l= g({\bf{y}}_{l-1},p_l) \, ,
\end{equation}
where the function $g$ is just the procedure described above to generate strings
from the CMN dynamics starting from ${\bf{x}}_{0}={\bf{y}}_{l-1}$ and ending at
the first occurence of symbol $p_l$. Thus, function $g$ is the recipe by which a
sequence of ASCII symbols are associated to the sequence of state vectors arising
from the evolution of the CMN. The autonomous evolution of the CMN system yields the string
\begin{equation}
\label{alpha}
 \alpha=g({\bf{x}}_0,p_1)\,\cdot \, g({\bf{y}}_{1},p_2)\,\cdot \, .\, .\,
.\,\cdot \, g({\bf{y}}_{l-1},p_l) \, \cdot \, .\, .\, .\, .
\end{equation}

The segmentation of string $\alpha$ by a finite string
$\rho=(p_1,p_2,\ldots,p_n)$ can be represented by a $n$-dimensional vector
${\bf{c}}(\rho)$ whose components are the natural numbers giving the lengths
$|\beta_l \,\cdot \, p_l|$. That is,
\begin{equation}
\label{cdrho} {\bf c}(\rho)=
(|g({\bf x}_0,p_1)| \, , \, |g({\bf y}_{1},p_2)|\, \ldots,
|g({\bf y}_{n-1},p_n)|) \, .
\end{equation}
Since the CMN can be iterated indefinitely, Eq. (\ref{cdrho}) just expresses the
segmentation ${\bf c}(\rho)$ for the first
$T=\sum_{i=1}^n \, |g({\bf y}_{i-1},p_i)|$
symbols of string $\alpha$. In the example given in Eq.(\ref{ejemp_s}), one gets
${\bf c}(\rho)=(4,4,6,5,4)$.

Segmentation ${\bf c}(\rho)$ in Eq. \ref{cdrho} provides the position of symbols
$(p_1,p_2,...,p_n)$ in string $\alpha$, and therefore it can be used as the
encryption of the plaintext $\rho =(p_1,p_2,...,p_n)$. That is, after a number of
$t=\sum_{i=1}^l \, |g({\bf y}_{i-1},p_i)|$ iterations from ${\bf x}_0$ the
plaintext symbol $p_l$ is generated by the CMN evolution. If the local dynamics
$f(x)$ is public, the secret key may
consist of the couplings $\epsilon _{ij}$ and the
initial condition ${\bf x}_0$. Once the couplings are
specified, the autonomous evolution of the CMN will generate a string $\alpha$
that depends only on ${\bf x}_0$. In other words, under autonomous evolution, 
if the matrix $E$ and ${\bf x}_0$ are used as a secret key, string $\alpha$
will always be the same for a fixed key. Therefore, a number of $n=|\rho|$
symbols of string $\alpha$ can be known if the plaintext $\rho$ and its
corresponding ciphertext ${\bf c}(\rho)$ are known. Unknown elements in between
the $n$ known symbols $(p_1,p_2,...,p_n)$ can be inferred by using new messages
encrypted with the same key, even when the new plaintexts are unavailable. 
In the
example given in Eq. (\ref{ejemp_s}), the word ``Rival" is encrypted as
${\bf c}(R,i,v,a,l)=(4,4,6,5,4)$; therefore, after 19 iterations and after 23
iterations the CMN generates the symbols ``a'' and ``l'', respectively. If another
word has an encryption ${\bf c}(\rho)=(4,4,4,4,3,4)$, it can be guessed that
$\rho=(R,i,t,u,a,l)$, and two new symbols of the string $\alpha$ can be inferred.

Note that this decoding method is possible because, for the autonomous 
evolution of the CMN,
the string $\alpha$ is unique for a given key. In order to avoid this limitation,
a non-autonomous CMN evolution can be used. A possibility is to make string
$\alpha$ dependent on the plaintext to be encrypted. We call this method Text
Dependent Encryption (TDE). An example of TDE is a CMN dynamics that is perturbed
each time that a symbol is encrypted. This perturbation could be, for example, a
change of sign in the states of the  maps $x_t^i$ each time that a symbol $p_l$
is encrypted. In this case, when the plaintext $\rho=(p_1,p_2,...,p_n)$ is
encrypted, the resulting CMN evolution string $\alpha$ can be expressed as
\begin{equation}
\label{alpha_ym}
\alpha(\rho)=g({\bf x}_o,p_1)\,\cdot \,
g(-{\bf y}_{1},p_2)\,\cdot \, .\, .\, .\,\cdot \,
g(-{\bf y}_{n-1},p_n) \, ,
\end{equation}
and the corresponding encryption is
\begin{equation}
\label{c_ym}
{\bf c}(\rho)= (|g({\bf x}_o,p_1)| \, , \,
|g(-{\bf y}_{1},p_2)|\, , \, .\, .\, .\, ,
|g(-{\bf y}_{n-1},p_n)|) \, .
\end{equation}

Note that the $l$-th segment $\beta_l \cdot p_l$ of the string $\alpha$ is
univocally determined by the previous $(l-1)$ symbols in the plaintext $\rho$.
Conversely, encryption ${\bf c}(\rho)$ in Eq. (\ref{c_ym}) allows to reproduce
the CMN dynamics generated during the encryption process, since ${\bf c}(\rho)$
indicates at which iteration steps the dynamics must be perturbed. Therefore,
both the string $\alpha(\rho)$ and the plaintext $\rho$ can be reconstructed if
the appropriate key is used (i.e. the appropriate CMN parameters and initial
condition). We call this encryption method TDE(*-1), to indicate that the CMN
vector state is multiplied by (-1) each time that a symbol is encrypted. The
encryption with the autonomous CMN evolution can be denoted by TDE(*+1). Other
operations can be used in the TDE method.

In principle, any aperiodic function $f(x)$ can be used as local map in the CMN
system, 
Eq. (\ref{cml}). As an example, we consider unbounded local chaotic dynamics
given by the logarithmic map \cite{Kawabe}
$f(x)=b + \ln |x|$. This map is chaotic, with no periodic windows, 
on the parameter interval $b \in (-1,1)$.
The unbounded character of the local functions places no restrictions on the 
range of parameters values of the 
CMN system that can be explored. 
For local parameter values about $b \approx 0.5$,
and the couplings randomly selected in the interval $|\epsilon_{ij}| < 0.1$, 
$\forall \; i,j$, all
the ASCII symbols in the set $Z_{128}$ are generated by the 7-dimensional CMN
with about the same probability of $1/128$, as can be seen in Fig. 1. 
This shows that all the
$2^7$ ``Cartesian quadrants" are visited by the state vector of the CMN in  about
$2^7$ iterations. Note also that the standard deviations of substrings lengths
are of the same order of magnitude than its average, which is typical of an
aperiodic string.
\begin{figure}[ht]
\centerline{\hbox{
\epsfig{file=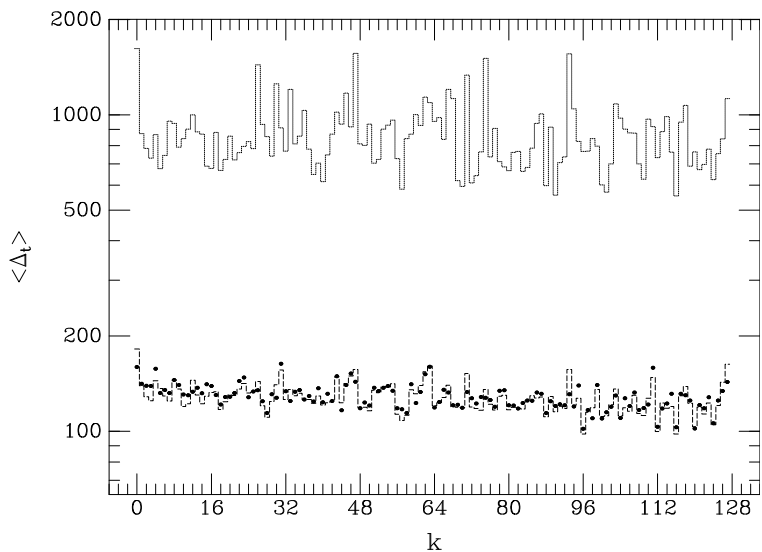,width=.40\textwidth,angle=-90,clip=}
}}
\caption{ Mean distance between successive occurrences of the same symbol $z_k$
in an autonomous string $\alpha$ of length $|\alpha|=50,000$ 
as a function of $k$, ($k=1,2, ... , 128$), for fixed $b=0.47$. 
When $\rho$ is a string
consisting of a repetition of the same symbol $z_k$ (i.e. $\rho=z_k,z_k,...\,
\equiv {\bar{z_k}})$, the dots give the average $<\Delta_t> \equiv <|\beta_l
\cdot z_k|>$ of the length of segments in ${\bf{c}}({\bar{z}}_k)$ (see Eq.
(\ref{cdrho})). The dashed curves show the standard deviation of the segment
lengths in ${\bf{c}}({\bar{z}}_k)$. The upper dotted curve displays the maximum
values of the segment lengths $\Delta_t({\bar{z_k}})$ in the $50,000$ iterations;
the minimum segment lengths lie between 1 and 4. }

\end{figure}

Another useful property of chaotic CMNs as encrypting schemes is their
sensitivity to initial conditions and/or couplings. The sensitivity to the
couplings can be measured by comparing two strings, $\alpha$ and $\alpha'$,
generated by two CMNs identical to each other, except by one element in the their
coupling matrices, $\epsilon_{ij}'=\epsilon_{ij} + \delta_{ij}$. Figure 2 shows
the mean number of iterations $<t_{diff}>$ at which the strings $\alpha$ and
$\alpha'$ start to differ, as a function of the size of the perturbation
$\delta_{ij}$. The various curves correspond to different truncations of the
CMN states after each iteration $t$. The truncation used consists of expressing
the real value of each component of the state vector ${\bf{x}_t}$ with a given
number of significative digits. The truncated state ${\bf{u}_t}$ is used to
calculate the state ${\bf{x}_{t+1}}$ at iteration $t+1$. That is,
${\bf x}_{t+1}= {\bf f}({\bf u}_t)+{\bf E} {\bf u}_t$. This truncation is
relevant since it can be used to make the numerical process equivalent in
computers with different precisions.
\begin{figure}[ht]
\centerline{\hbox{
\epsfig{file=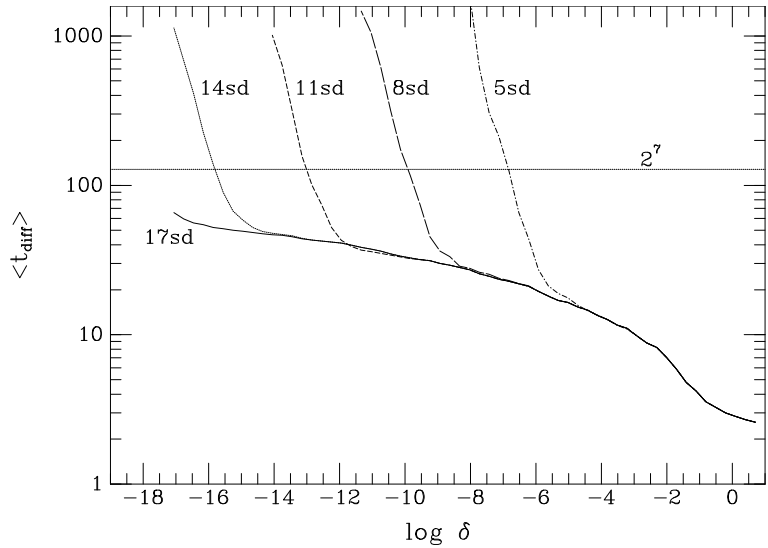,width=.40\textwidth,angle=-90,clip=}
}}
\caption{ Mean number of iterations $<t_{diff}>$ as a function of the size of the
perturbation $\delta$, for $b=0.47$. The various curves correspond to different
truncations of the CMN states after each iteration $t$. The value $<t_{diff}>$
is obtained averaging the $7 \times 7$ results $t_{diff}(\delta_{ij}=\delta)$
for $i \, , j=1,2, ... ,7$. The labels indicate the number of significative
digits of the components of ${\bf{u}_t}$ (i.e., 17sd indicates 17 significative
digits).}
\end{figure}

Since the typical number of iterations to find a given symbol is about $2^N$, we
measure the encrypting sensibility $\delta_{cri}$ of the CMN as the value of
$\delta$ for which $<t_{diff}> = 2^N $. Note that $\delta_{cri}$
is a very small value (ej. $ \sim 10^{-12}$ for a 10 significative digits
truncation).

For a fixed value of parameter $b$ and $N=7$, the maximum encrypting key consist
of $7 \times 7$ coupling strengths and $7$ initial conditions $x_0^i$. As shown
in Fig. 2, a change of $\delta=10^{-10}$ in one of the coupling strengths is more
than enough to modify the string $\alpha$ after $t \approx 40$. Therefore, there are
more than $10^{\delta^{-1} \times N \times (N+1)} \sim 10^{560}$ possible keys.
Obviously, among all of these possibilities there are groups of keys that
produce strings $\alpha$ that are identical to each other up to $t_{diff} \gg 40$,
but the probability of finding two of such keys is very small ($\sim 2^{-N
\times t_{diff}}$). In general, such large number of possible keys is unnecessary,
and in practice the key can be reduced by using a set of random number seeds that
are used to generate the $7 \times 7$ coupling strengths and the $7$ initial
conditions $x_0^i$. Alternatively, the system size $N$ can be reduced in order to
decrease the number of possible keys and to increase the encrypting speed.

As an example, for $b=0.5$, $x_0^i=1.0+0.1 i$, and $\epsilon_{ij}=0.01(i-j/2)$
($i,j=1,...,7$), the encryption of the text ``Rival ritual'' would be:

a) using the autonomous CMN evolution (Eq. (\ref{cdrho})),
\begin{center}
128 44 18 530 33 505 7 206 97 95 8 170
\end{center}

b) using the encryption method TDE(*-1) (Eq. (\ref{c_ym})),
\begin{center}
128 387 64 34 36 3 96 297 146 26 78 3
\end{center}

c) using the encryption method TDE(*-1) (Eq. (\ref{c_ym})), but adding the small
quantity $10^{-10}$ to the coupling weight $\epsilon_{3 \,5}$,

\begin{center}
425 176 20 156 8 85 234 43 32 87 224 80
\end{center}

If we try to recover the plaintext using the encryption method TDE(*-1)
in (b), but using the coupling $\epsilon_{3 \, 5}$ altered by the amount
$\delta_{3 \,5}=10^{-10}$, the resulting decoded text is:

\begin{center}
0$<$3w 7h\$$|$lR"
\end{center}

In the above example, the FORTRAN internal function ICHAR have been used to assign
a binary 7-digit number to each of the ASCII symbols in the set $Z_{128}$.

\begin{figure}[ht]
\centerline{\hbox{
\epsfig{file=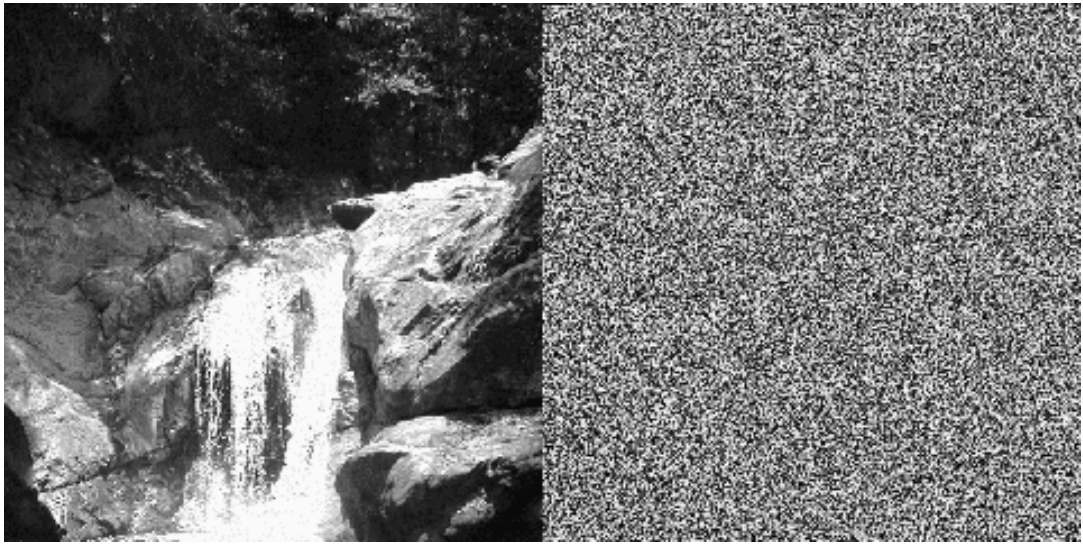,width=.80\textwidth,angle=0,clip=}
}}
\caption{The BMP image on the left has been encrypted assuming $b=0.5$,
$x_0^i=1.0+0.1 i$ and $\epsilon_{ij}=0.01(i-j/2)$ ($i,j=1,...,8$). On the right,
the corresponding decoded image is shown when a slightly wrong key is used: we have added
$10^{-10}$ to $\epsilon_{3 \, 5}$.
}
\label{fig3}
\end{figure}

Notice that using $N=8$, the CMN dynamics can be employed to
generate strings with elements in a pallete of 256 gray tones and therefore to
encrypt images byte by byte, as shown in Fig. \ref{fig3}.

In conclusion, we have shown how a CMN can be used as a string generator
and as an encrypting system. 
The ciphertext contains the
information on how the CMN must be evolved and how to select the plaintex symbols
from the string associated to the CMN's evolution. 
The secret key consists of the coupling matrix ${\bf E}$ and the initial
conditions. The number of parameters involved in the secret key and the
high sentivity of the generated strings to small 
perturbations of any of those parameters make the CMN encrypting 
scheme difficult to break.
This confers an advantage to this scheme, in terms of security, in comparison 
to communication procedures based solely 
on one chaotic dynamics. 
The use of several coupled dynamics instead of
just one allows the transmition of entire sequences of 
the plaintext at a time.  
The notation introduced allows to place 
the proposed encrypting method in a wide context.
The implementation of
variations of the method is straightforward.
The  examples presented here 
show the encrypting performance for a network of $7$ coupled logarithmic maps, 
however the method can be applied with networks of any size.
Finally, we note that the CMN parameters determine both the probability of
occurrence of symbols and the transition probability $p(z_j|z_i)$ of
observing the symbol $z_i$ followed by symbol $z_j$ in the string.
Therefore, it is possible in principle to select the CMN parameters in
order to enhance or to inhibit some symbols and transitions.
Since this is equivalent to select grammatical rules, the CMN as string
generators can be of interest in the development of language models.

\section*{ Acknowledgments}
M.G.C and A.P acknowledge support from Consejo de Desarrollo Cient\'{\i}fico,
Human\'{\i}stico y Tecnol\'ogico of Universidad de Los Andes, M\'erida, Venezuela. The
authors thank Professor R. Pino for useful comments.

\end{document}